\documentclass{mem}
\usepackage{natbib}\usepackage{txfonts}\usepackage{balance}
\usepackage{graphicx}
\usepackage{txfonts}
\usepackage[a4paper]{hyperref}
\idline{79}{1}
\begin{document}
\def\teff{$T\rm_{eff }$}
\def\kms{$\mathrm {km s}^{-1}$}

\title{
The Bologna Open Clusters Chemical Evolution project
(in short: BOCCE)
}

   \subtitle{}

\author{
A. \,Bragaglia 
}

  \offprints{A. Bragaglia}

\institute{
INAF --
Osservatorio Astronomico di Bologna, via Ranzani 1, 40127 Bologna, Italy
\email{angela.bragaglia@oabo.inaf.it}
}

\authorrunning{Bragaglia }

\titlerunning{BOCCE project}

\abstract{
I present here our project, the Bologna Open Clusters Chemical Evolution (BOCCE)
project, aimed at using Open Clusters as tracers of the disk properties and
their evolution with time. We are collecting and homogeneously analyzing data, both photometric and spectroscopic,
on a large sample of open clusters, representative of the old cluster population, and
I show here results obtained on a subset of our clusters.
}
\maketitle{}

\section{Our project}

Galactic open clusters (OCs) are considered very good tracers of  
the disk' properties: they are seen over the
whole disk, cover the entire age interval of the disk and trace its chemical
abundances both at present time and in the past  \citep[e.g.,][]{friel95}. 

One of the subjects where there are several advantages in using OCs instead of isolated, field stars, is the study of
the disk metallicity distribution and
its possible evolution with time. As a matter of fact, 
the distances and ages of OCs can be measured with higher precision up to large distances,
and their ages span a much larger interval than e.g., B stars, Cepheids or Planetary Nebulae, 
other widely used tracers.

OCs are also useful tests for stellar models and are complementary to older,
metal-poorer globular clusters.

If we want to study the history of the disk, we have to obtain
information on a large and significant number of OCs; of course, old OCs must
be conspicuously present in this sample. 
With our program BOCCE, that stands for Bologna Open Cluster Chemical
Evolution project, we are building and homogeneously analyzing such a sample; for a  
a detailed description of its goals and a summary of results
on the first part of the photometric work, see  \citet{bt06}. 

Very briefly, we employ 
i) deep, precise photometry to derive ages, distances and reddeninigs (and a first indication of the metallicity)
using the comparison of observed and synthetic colour-magnitude diagrams (CMDs, e.g.
\citealt{bt06}); 
ii) medium resolution spectra to derive radial velocities and crucial information on
membership \citep[e.g.,][]{vale}; 
iii) and high resolution spetra to derive the metallicity and the detailed abundances 
\citep[e.g.,][]{cbgt04,cbgt05}.

\begin{figure}[t!]
\resizebox{\hsize}{!}{\includegraphics[clip=true]{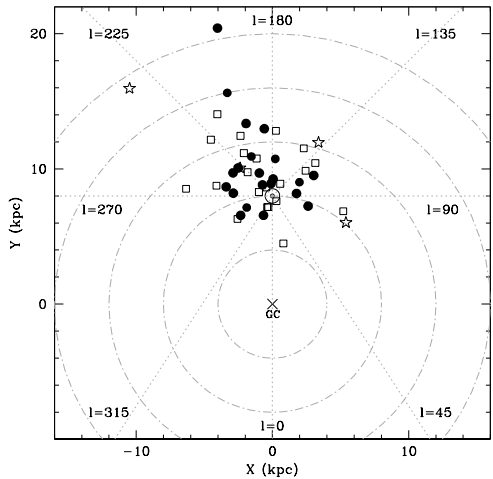}}
\caption{\footnotesize
Positions of the clusters in the BOCCE sample; filled symbols indicate objects for which
the photometric analysis is completed, stars indicate work in progress, and empty squares
clusters acquired but not studied yet.}
\label{fig1}
\end{figure}

We try to cover all disk positions both in distance and direction, as shown in Fig.~\ref{fig1}, where  we plot all the OCs for which photometry has been acquired. We concentrated on old
clusters and have already published results for 16 OCs older than about 1 Gyr, while a few more
are expected soon. This number represents a fair fraction of the total number of similar, known
clusters: in the most recent catalogue by  \cite{dias} there are about 120 OCs older than 1 
Gyr, out of the more than 1700 objects present.

We have already obtained large amounts of data, but the analysis has been
completed only for  part of them.
Furthermore, we also plan to increase our sample including interesting 
 clusters  from the archives or from collaborations, and
homogenizing their analysis to our system. An example of the latter will be
the  distant OCs observed with FLAMES/UVES (PI S. Randich) in the 
anticenter direction (see e.g., \citealt{paola} and Sestito et al. this conference).

\subsection{Photometric data}
We have alredy published results for 20 clusters \citep{bt06,n3960,be17,tbc07},  which
represent about one half of our 
sample and that cover the age range from about 0.1 to 9 Gyr. We are presently adding other ones, like Be~20 and Be~66 \citep{gloria}, or NGC~6791.

The homogeneous determination of ages, distances, reddening and a first indication of metal abundance is our
main result. They are derived  using the photometric data and synthetic CMDs based
on stellar evolutionary tracks and taking into account photometric errors and completeness, and 
the presence of a fraction of binaries.  
We always use the same three sets of tracks: the FRANEC, without overshooting \citep{franec},
the old Padova ones,  with classical overshooting \citep[e.g.,][]{pd}, and the FST ones, which use  the Full Spectrum Turbulence approach
described by \cite{fst}. 

Once we have completed our homogeneous analysis for a large enough number of targets, 
we can compare the clusters' properties on a
common scale, and so derive information on the Galactic disk; and
we may compare the influence of different assumptions on the
measured parameters, and test stellar models. 
Since we derive all parameters with the three
sets,  we have a good estimate  of the systematics involved.

\begin{figure*}[]
\resizebox{\hsize}{!}{\includegraphics[clip=true]{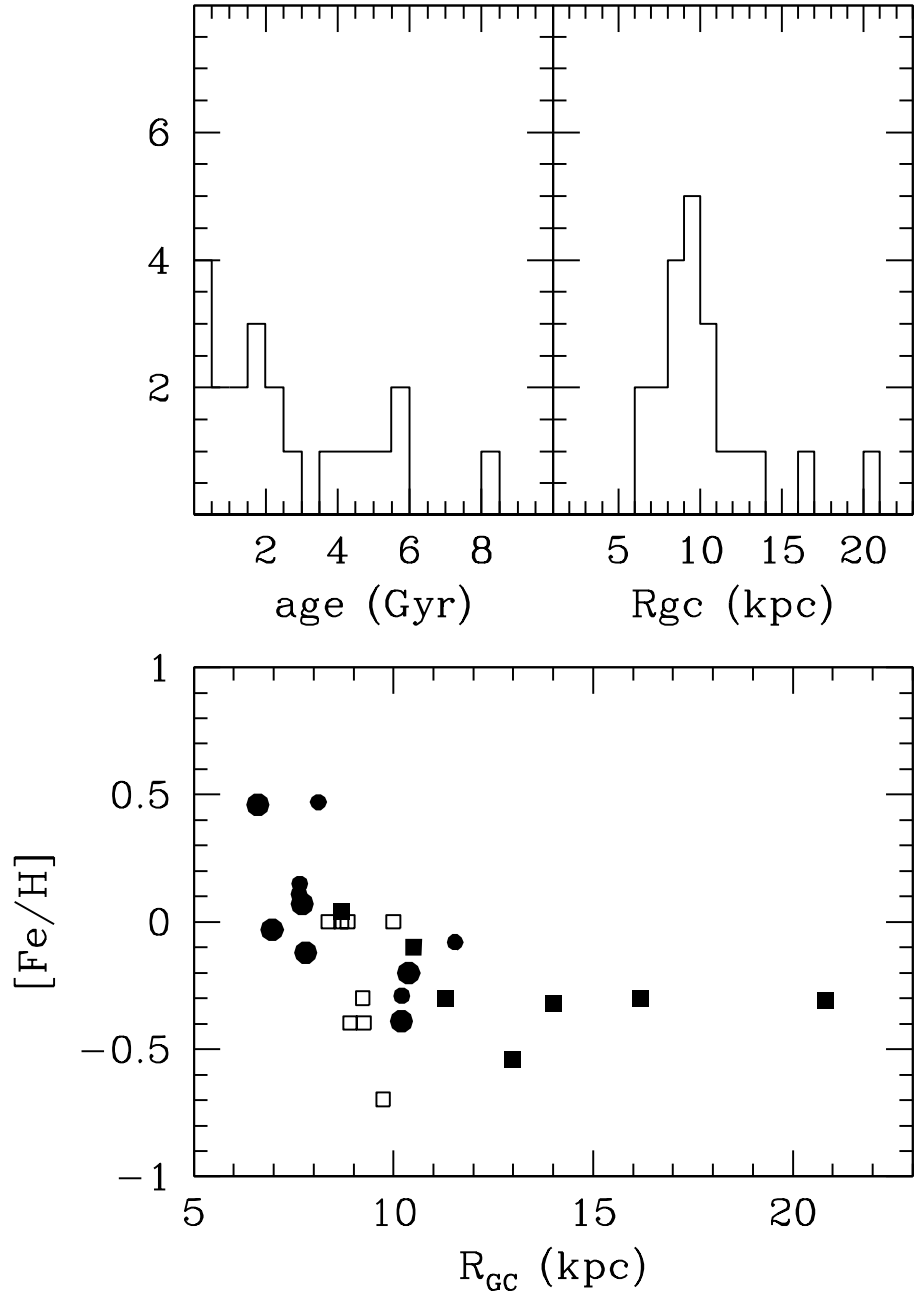}}
\caption{\footnotesize
Upper panels: histograms of the ages and distances fron the Galactic center for BOCCE
clusters. Lower panel: the radial abundance distribution. Different symbols
indicate the degree of integration in the BOCCE sample: large filled circles:
all parameters obtained in the BOCCE system; smaller filled circles: [Fe/H]
obtained in BOCCE, R$_{GC}$ from literature; filled squares [Fe/H] from
literature, R$_{GC}$ in BOCCE; squares: R$_{GC}$ in BOCCE, but [Fe/H] 
from literature (filled) or from the BOCCE photometry,
i.e., from the Z of the evolutionary tracks (open), respectively.  
}
\label{fig2}
\end{figure*}

We have a few very interesting objects in our sample.
In particular, we have observed Berkeley 29  \citep{be29}, the farthest known cluster, very
important to define any metallicity gradient, and Berkeley 17 \citep{be17}, which is perhaps the oldest known open cluster, with an age similar to the one of
the youngest globular clusters.
We are presently working on the photometry of NGC~6791, a rather peculiar
cluster, that could be the oldest (but see Be~17) and metal-richest (but see NGC~6253) in our Galaxy.
We use the deep and precise data obtained with the CFHT \citep{jason07}, taking also into account
information
on membership from radial velocities, very useful to better define its red giant branch. We have just started the simulations of its CMDs, but its very high metallicity is an obstacle, since of the models 
that we homogeneously use, only the Padova tracks reach the required  Z. We need
new, more metal-rich extension of the two other sets (in preparation).

\subsection{Spectroscopic data}
We have already analyzed the high-resolution spectra of about 10
clusters \citep{n6819, cbgt04,cbgt05,gbct06,cbg07}, which span the metallicity from [Fe/H] $\simeq-0.5$
to [Fe/H] $\simeq+0.5$ dex.
We already  have data on a few more objects, 
and have recently obtained observing time to complete the spectroscopic part of other OCs for which the photometry has been presented.
Further data  will be added, from the archives and
from a companion
program, for other interesting clusters.

Also in this case, our goal is to reach the highest possible precision and homogeneity.
To ensure this,  we always use the same model grids to derive abundances, the same
line lists,  $gf$'s, solar reference abundances, and the same method of
measurement of EWs  or synthetic spectra. 

All our spectra have been obtained up to now with SARG@TNG,  FEROS@1.5m ESO, and
UVES@VLT; they have a resolution $R\sim30000-50000$. 
Our strategy is to obtain  spectra of a few stars (3-5) in each
cluster, chosen among confirmed members by previous radial velocity or proper
motion studies. We usually  concentrate on red clump giants, since they are  the best
compromise between the bright luminosity necessary to reach very good S/N even at
high  resolution and temperatures not too cold to be a problem for the analysis
of line-crowded spectra.

One of the motivation of our project is to determine the metallicity distribution in the Galactic disk.
We have not reached  our goal yet, but first results can be seen in Fig.~\ref{fig2}.
The upper panels show the distributions in age and distance from the Galactic center
of all cluster already photometrically analyzed. The lower panel is a representation of
the radial metallicity distribution. However, 
this plot is not completely based on BOCCE, because the analysis is done completely by our group
on a common scale only for part of the clusters.   At the moment we cannot yet
derive a self-consistent picture of the radial
metallicity distribution or of its possible evolution with time: we need to reach full
homogeneity.

One may wonder whether
OCs really are good tracers of  
the Galactic disk. When we compare their abundances (or better the run 
of elemental ratios with [Fe/H]) with those of field stars, the answer
seems to be positive. In general the elemental ratios follow the same
pattern for OCs and field stars. There are a few exceptions, like Na; this
could mean for instance that we are not taking into
account well the non-LTE effects for Na, or maybe that there are actual differences
between giants (usually used to study clusters) and dwarfs (usually selected
in field samples). This is  a complicate subject and needs dedicate studies.

An example of "good" behavior comes from the $\alpha$-elements: they share
the same run with metallicity for field and cluster stars, in all the metallicity range
covered by OCs. Furthermore, the [$\alpha$/Fe] ratios do not show any dependence
from the cluster age, in our sample or in literature ones.
There is an indication of a slight trend for [$\alpha$/Fe] values to increase with 
Galactocentric distance; however, this is based on literature abundances, 
and seems not confirmed by other studies (see Sestito et al., this conference).
Further discussion is postponed until we have homogeneously analyzed all the BOCCE
sample.

\section{Summary}

To briefly summarize : \begin{itemize}

\item
we are studying (old) open clusters as tracers of the
disk properties;

\item
to this end we are deriving ages, distances, reddenings, metallicities and detailed 
abundances for a large sample of old OCs (about 20 already analyzed photometrically,
and about 10 spectroscopically, up to now);

\item
in doing so, we try to maintain the maximum homogeneity of methodology, to reach really sound and significant results; 

\item
if we compare them to field stars, the open clusters do indeed appear to be good tracers of the general disk abundances, confirming that we may use them to trace the disk properties;
 
\item
abundances of open and globular clusters and field stars  studied 
by our group will be on a common scale, ensuring meaningful comparisons between
different stellar populations;
\item
finally, our sample can be used to test stellar evoutionary models of different ages
and metallicities, being complementary to the older, metal-poorer globular clusters.

\end{itemize}

\begin{acknowledgements}
People currently collaborating in this project are: M. Tosi, E. Carretta, M. Cignoni
(INAF-Oss. Astr. Bologna), R.G., Rratton, E.V. Held (INAF-Oss. Astr. Padova),
G. Andreuzzi, L. Di Fabrizio (INAF-Fundaci\'on Galilei), J. Kalirai (UCO-Lick),
and G. Marconi (ESO). This study has been made possible by generous allocation
of time at Italian telescopes (Loiano and TNG), at the CFHT, and at ESO telescopes, 
both in La Silla and Paranal. This work has been greatly helped by the WEBDA (located at 
http://www.univie.ac.at/webda/).

\end{acknowledgements}

\bibliographystyle{aa}

\end{document}